\documentstyle[12pt,aps,prb]{revtex}
\catcode`\"=\active \let"=\" \let\3=\ss
\renewcommand{\d}{\displaystyle}
\newcommand{\nnumb}{\begin{displaymath}\hspace{-1cm}
 \begin{array}{rcl}}\newcommand{\nnume}{\end{array}\end{displaymath}}
\newcommand{\Kr}[1]{\left( #1\right)}
\newcommand{\Ke}[1]{\left[ #1\right]}
\newcommand{\Ks}[1]{\left\langle #1\right\rangle}

\newcommand{\gapprox}{\mbox{\raisebox{-4pt}{$\,\buildrel>\over\sim\,$}}}
\newcommand{\D}{\mbox{$\cal D$}}
\newcommand{\K}{\mbox{$\cal K$}}
\newcommand{\F}{\mbox{$\cal F$}}
\def\fcap#1{\rightskip1cm\noindent\small #1\normalsize\rightskip0cm}
\begin{document}
\title{Non-linear current through a barrier in 1D wires with
finite-range interactions}

\author{Margit Steiner$^*$ and Wolfgang H"ausler$^{**}$}

\address{I.~Institut f"ur Theoretische Physik der Universit"at Hamburg,\\
Jungiusstr.~9, D-20355 Hamburg, Germany}

\maketitle

\begin{abstract}
The transport properties of a tunnel barrier in a one-dimen\-sional
wire are investigated at finite voltages and temperatures.
We generalize the Luttinger model to account for finite ranges
of the interaction. This leads to deviations from the power law
behaviour first derived by Kane and Fisher \cite{kanefisher}.
At high energies the influence of the interaction disappears and
the Coulomb blockade is suppressed. The crossover in the
voltage or in the temperature dependence can provide a direct
measure for the range of the interaction.

\end{abstract}

\noindent Keywords~: electron-electron interactions,
electronic transport, tunnelling.
\footnotetext{Present addresses~: $^*$ Imperial College, Math.
Phys. Department, 180, Queen's Gate, London SW7 2BZ, UK\\
$^{**}$ Theoretical Physics Institute, 116
Church Str. SE Minneapolis MN 55455, U.S.A.}

\noindent Since the discovery of the vanishing transmittivity of
a tunnel barrier in a one-dimensional (1D) wire due to the
repulsive electron--electron interaction \cite{kanefisher} new
interest has emerged in the transport properties of 1D electron
systems\cite
{furusaki93b,sassetti94,matveev93a,glazman94a,gogolin94a,ludwig95a,weiss95,egger96,weiss96}.
Indeed, the influence of the electron correlations shows up
strikingly in non-linear current voltage relations
which are investigated experimentally in narrow, semiconducting
wires \cite{tarucha95}. Local interactions, $\:v(x-x')=
v_0\delta(x-x')\:$ are described within the Luttinger model for
which the power-law
\cite{kanefisher,furusaki93b,sassetti94,matveev93a,glazman94a}
\begin{equation}\label{iv}
I(V)=\frac{\omega_c}{eR_{\rm T}}\Kr{\frac{eV}{\omega_c}}^{2/g-1}
\end{equation}
has been predicted for the current-voltage relation through a
tunnel barrier with tunnel resistance $\:R_{\rm T}\:$, at zero
temperature $\:T=0\:$. A similar behaviour has been found for
the linear conductance $\:\sim T^{2/g-2}\:$ as a function of
temperature \cite{kanefisher,furusaki93b,glazman94a}. The
exponent depends on the strength $\:v_0\:$ of the interaction,
$\:g\equiv(1+v_0/\pi v_{\rm F})^{-1/2}\:$ ($\:v_{\rm F}\:$:
Fermi velocity). Repulsion corresponds to $\:g<1\:$.

The Luttinger model limits the energies to values well below the
upper cut-off $\:\omega_c\:$ which can be identified with the
Fermi energy and can be small in semiconducting devices at low
electron densities. At larger voltages, Eq.~(\ref{iv}) formally
describes currents that exceed even the value $\:V/R_{\rm T}\:$
for non-interacting electrons. This indicates that some new
energy or length scale must become important at higher energies.
Here we shall identify the finite range of the
$\:e-e$--interaction to cause an asymptotic approach of the
current towards $\:V/R_{\rm T}\:$ at large voltages. In the
case of very strong interaction the cross-over can occur in an
oscillatory manner while for more realistic interaction
strengths the current stays below $\:V/R_{\rm T}\:$ for all
voltages. The low energy, and hence long wave length properties,
are well described within the Luttinger model where the range of
the interaction $\:1/\alpha\:$ is assumed to be shorter than
even the inter-electron spacing $\:a\equiv\pi v_{\rm
F}/2\omega_c\:$ for spinless electrons. Finite voltages,
however, introduce a wave length $\:1/\Delta k = v_{\rm F}/eV\:$
on which eventually a finite value for $\:1/\alpha\:$ can be
experienced. Large voltages change the momenta at the Fermi
points by $\:|\Delta k|\:$ that exceeds the scale $\:\alpha\:$
on which the Fourier components of the interaction vanish,
$\:\hat{v}(\Delta k>\alpha)\approx 0\:$ so that $\:V>\alpha
v_{\rm F}/e\:$ {\em suppress the effect of interactions.}

Accordingly, at high temperatures $\:T>\alpha v_{\rm F}\:$,
$\:I(V)\:$ becomes independent of $\:T\:$ so that the
differential conductance approaches the constant $\:1/R_{\rm
T}\:$, like in the non-interacting case (we do not consider the
effect of lattice vibrations here \cite{brandes96}). Finite
temperatures may even reduce the current. Both the voltage and
the temperature dependencies of the current show the important
common feature of a crossover which in principle allows to
extract the range $\:1/\alpha\:$ of the interaction.

The importance of finite range interactions has been found
already in zero dimensional systems, quantum dots, where
`crystallisation' of the charge density distribution can occur
leading to qualitative changes in the low energy excitation
spectra \cite{hausler95,weinmann95a,jefferson96} as
compared to what is expected for a contact interaction
\cite{jauregui93}.

For convenience we concentrate on the interaction \cite{cuniberti}
\begin{equation}\label{interaction}
v(x-x')=v_0\frac{\alpha}{2}{\rm e}^{-\alpha|x-x'|}\quad.
\end{equation}
In the presence of metallic gates close to the 1D channel a more
realistic form would vary $\:\sim|x-x'|^{-3}\:$ at large
distances, however, we expect qualitatively the same results for
the latter finite range interaction as for (\ref{interaction}),
cf.\ below. The non-Fermi liquid behaviour of the charge
excitations in a 1D wire is expressed most conveniently by the
Hamiltonian
\cite{egger96,haldane,schulz93b,fisherglazman}
\begin{eqnarray}\label{hw}
H_{\rm w}&=&\d\frac{v_{\rm F}}{2}\int{\rm d}x\;[(\Pi(x))^2+
(\partial_x\theta(x))^2]+\\[3ex]
&&\d\frac{1}{2\pi}\int{\rm d}x\;{\rm d}x'\;
(\partial_x\theta(x))v(x-x')(\partial_{x'}\theta(x'))\nonumber
\end{eqnarray}
where the Fermi-fields are expressed \cite{kanefisher,haldane}
through Bose fields, $\:\Pi(x)\equiv\partial_x\phi(x)\:$ and
$\:\theta(x)\:$, with $\:[\phi(x)\,,\,\theta(x')]=-({\rm
i}/2)\mbox{\rm sgn}(x-x')\:$. The spatial derivative,
$\:\partial_x\theta\:$, measures the fluctuations of the charge
density, and the time derivative, $\:\partial_t\theta\:$, is
proportional to the current.

Here, we account for the dispersion relation of the
charged modes in the wire, as it can be obtained from (\ref{hw})
by spatial Fourier transform
\begin{equation}\label{dispersion}
\omega(k)=v_{\rm F}|k|\sqrt{1+\frac{\hat{v}(k)}{\pi v_{\rm F}}}\quad.
\end{equation}
The Fourier transform of the interaction potential (\ref{interaction})
\begin{equation}\label{fourier}
\hat{v}(k)=v_0\frac{\alpha^2}{k^2+\alpha^2}
\end{equation}
is constant $\:\hat{v}(k)=v_0\:$ in the limit
$\:\alpha\to\infty\:$ of the Luttinger liquid used in previous
calculations where it merely renormalizes the sound velocity
$\:\to v_{\rm F}/g\:$.

The tunnel barrier can be described \cite{kanefisher} by
\begin{equation}\label{barrier}
H_{\rm b}=U_{\rm b}\Ke{1-\cos(2\sqrt{\pi}\theta(x=0))}\quad.
\end{equation}
Furthermore, we assume an electrostatic potential
$\:(V/2)\mbox{sgn}(x)\:$ dropping discontinuously at the
location $\:x=0\:$ of the tunnel barrier,
\[
H_V=\frac{eV}{\sqrt{\pi}}\theta(x=0)\quad,
\]
as in \cite{kanefisher,furusaki93b,sassetti94,ludwig95a,weiss95,weiss96}.
In the limit of weak tunnelling it has been demonstrated \cite{egger96}
that the selfconsistently adjusted chemical potential indeed
varies most pronouncedly close to $\:x=0\:$.

The {\sc dc}--current
\begin{equation}\label{current}
I=-\frac{e}{\sqrt{\pi}}\Ks{\partial_t q(t)}
\end{equation}
can be expressed in terms of the field $\:\theta\:$ at $\:x=0\:$,
\[
q(t)\equiv\theta(t,x=0)
\]
where both, the quantum average and the dynamics refer to the
full Hamiltonian $\:H_{\rm w}+H_{\rm b}+H_V\:$.

Since $\:H_{\rm w}\:$ is purely quadratic in $\:\theta(x)\:$ all
of the contributions away from the impurity $\:x\ne 0\:$ can be
integrated out to obtain the reduced dynamics for $\:q(t)\:$.
We are interested in the probability for transitions of $\:q\:$
from a value $\:\theta_{\rm i}\:$ to $\:\theta_{\rm f}\:$ during
the time $\:t\:$ which can be expressed as a double integral
\begin{equation}\label{doublepath}
\int\limits_{\theta_{\rm i}}^{\theta_{\rm f}}\;
\D q\int\limits_{\theta_{\rm f}}^{\theta_{\rm i}}\;
\D q'\;\exp\Kr{{\rm i}S[q]}\exp\Kr{-{\rm i}S[q']}\F[q,q']
\end{equation}
over paths $\:q\:$ and $\:q'\:$ with endpoints
$\:q(0)=q'(t)=\theta_{\rm i}\:$ and $\:q(t)=q'(0)=\theta_{\rm
f}\:$. The action $\:S[q]\:$ contains all contributions to the
Hamiltonian at $\:x=0\:$ while the influence of the bulk modes,
$\:x\ne 0\:$, is exactly accounted for in the functional \cite{feynvernon63}
\begin{eqnarray}\label{influenz}
\F[q,q']&=&\d\exp-\int_0^t{\rm d}t'\;\int_0^t{\rm d}t''\;
(\dot{q}(t')-\dot{q}'(t'))\times\nonumber\\[3ex]
&&\d (w(t'-t'')\dot{q}(t'')-w^*(t'-t'')\dot{q}'(t''))
\end{eqnarray}
where
\begin{equation}\label{instint}
w(t)=\int_0^{\infty}{\rm d}\omega\;\frac{J(\omega)}{\omega^2}
\Ke{(1-\cos\omega t)\coth\frac{\beta\omega}{2}+{\rm i}\sin\omega t}\quad,
\end{equation}
and $\:1/\beta\:$ is the temperature. It will be important for
the following that the non-linear dispersion (\ref{dispersion})
of the bulk modes, shown in the inset of Figure~1, cause a
non-Ohmic dissipative influence. The function $\:J(\omega)\:$
includes all of the details of the environmental modes in their
efficiency to damp the frequency $\:\omega\:$ of the motion of
$\:\theta(x=0,t)\:$.

Within the Feynman--Vernon technique \cite{feynvernon63} the quantum
state for $\:q\:$ is assumed to be initially known (e.g.\
$\:\theta_{\rm i}=q(0)=0\:$) before the exact time
evolution is switched on. With time $\:t\:$,
$\:q\:$ acquires probability $\:P_m(t)\:$ to assume the value
$\:\theta_{\rm f}=m\sqrt{\pi}\:$ (cf.\ (\ref{doublepath})) where
$\:m\:$ elementary charges have been transferred through the
barrier. The long time behaviour of the probability density
distribution defines the stationary {\sc dc}--current
\[
I=-e\sum_m\:m\:\lim_{t\to\infty}\partial_tP_m(t)\quad,
\]
according to (\ref{current}), assuming ergodicity for the whole
system.

\addtocounter{footnote}{1}
For large $\:U_{\rm b}\:$ the potential (\ref{barrier}) has deep
minima at $\:\theta=m\sqrt{\pi}\:$ so that integer $\:m\:$
contribute mainly to the saddle points of the action $\:S[q]\:$
in (\ref{doublepath}). In this limit the charge is transferred
in integer units. Step like instantons dominate the path
integral (\ref{doublepath}) \cite{legett87,weiss91} for the low
current properties, each instanton contributing with a factor
$\:\pm{\rm i}\Delta/2\:$ proportional to the tunnelling
amplitude\footnote{The value of $\:\Delta\:$ can be related to
$\:U_{\rm b}\:$, cf.\ \cite{weiss96}. Through the one instanton
action, $\:\Delta\:$ depends in principle also on $\:\alpha\:$.}.

The influence functional $\:\F[q,q']\:$ (\ref{influenz})
introduces a temperature dependent coupling
$\:w(t_i-t_j)\:$ between instantons centred at times $\:t_i\:$
and $\:t_j\:$ so that the sum over all possible instanton
configurations in general cannot be performed analytically.
For a barrier of low transmittivity the most important
configurations are instanton -- anti-instanton pairs that
contribute in order $\:\Delta^2\:$. This leads to an
expression for the current \cite{sassetti94}
\begin{equation}\label{iv-final}
I(V)=e\frac{\Delta^2}{4}(1-{\rm e}^{-\beta eV})\int{\rm d}t\;
{\rm e}^{{\rm i}eVt}{\rm e}^{-w(t)}\quad,
\end{equation}
when the detailed balance property $\:\partial_tP_{-1}={\rm
e}^{\beta eV}\partial_tP_{+1}\:$ is used \cite{weiss93}.

For the case of Ohmic dissipation, $\:J(\omega)\propto\omega\:$,
corresponding to a contact interaction, considerable progress has
been made. To order $\:\Delta^2\:$ the current has been
calculated in \cite{sassetti94} and to any order for weak interaction
$\:1-g\ll 1\:$ in \cite{matveev}. Recently, the extension to
arbitrary interaction strength has been achieved using conformal
field theory techniques \cite{ludwig95a} and by systematically
exploiting the duality symmetry between low and high
transmittivities \cite{weiss96}.

\begin{figure}
\vspace*{7.6cm}\includegraphics{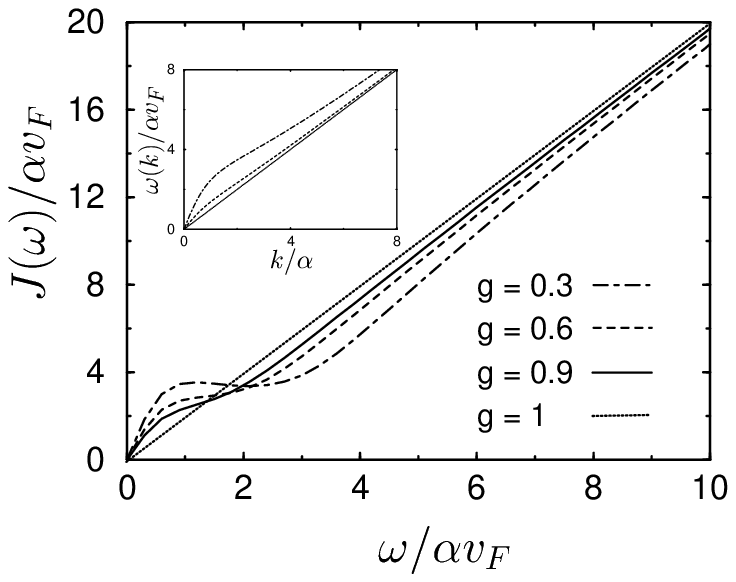}
\vspace{0.5cm}
\fcap{Figure~1: Effective density $\:J(\omega)\:$ of charged
modes in the wire that damp the motion of $\:\theta(x=0,t)\:$ at
the frequency $\:\omega\:$, according to Eq.~(\ref{analytical})
for different $\:g\:$. $\:J(\omega)\:$ is the crucial ingredient for the
non-linear current. The inset shows the dispersion relation
$\:\omega(k)\:$ according to Eq.~(\ref{dispersion}). Natural
units for wave vectors and frequencies are $\:\alpha\:$ and
$\:\alpha v_{\rm F}\:$, respectively.}
\end{figure}

How the electron--electron interaction influences the transport
properties is determined by $\:J(\omega)\:$ (cf.\
(\ref{instint},\ref{iv-final})). Its relationship to the bulk
mode dispersion $\:\omega(k)\:$ can most easily be deduced from
the partition function of the wire (\ref{hw})
\begin{equation}\label{singlepath}
Z={\rm Tr}\:{\rm e}^{-\beta H_{\rm w}}=
\int\D[\theta(x,\tau)]\;{\rm e}^{-S_{\rm w}[\theta]}=
\int\D[\hat{\theta}(k,\tau)]\;{\rm e}^{-S_{\rm w}[\hat{\theta}]}
\end{equation}
where
\[
S_{\rm w}[\hat{\theta}(k,\tau)]=\frac{1}{2v_{\rm F}}
\int\frac{{\rm d}k}{2\pi}\int_0^{\beta}{\rm d}\tau\;\hat{\theta}(-k,\tau)
\Kr{-\partial_{\tau}^2+\omega^2(k)}\hat{\theta}(k,\tau)
\]
with $\:\hat{\theta}(k,\tau)=\int{\rm d}x\;\theta(x,\tau){\rm
e}^{{\rm i}kx}\:$.

The modes $\:\theta(x\ne 0,\tau)\:$ act as a harmonic thermal
environment on the mode of interest, $\:q(\tau)\equiv\theta(x=0,\tau)\:$,
\begin{equation}\label{decomp}
Z=\int\D[q(\tau)]\int\D[\theta(x\ne 0,\tau)]\;
{\rm e}^{-S_{\rm w}[\theta]}\equiv\int\D[q]\;\varrho[q]\quad.
\end{equation}
The functional
\begin{equation}\label{redd}
\varrho[q]\propto\exp-\int_0^{\beta}{\rm d}\tau\;\int_0^{\beta}{\rm d}\tau'\;
q(\tau)\hat{\K}(\tau-\tau')q(\tau')
\end{equation}
for the reduced density contains the retarding effects, described
by the Kernel \cite{egger95}
\begin{equation}\label{kdef}
\K(\omega_n)=\int_0^{\beta}{\rm d}\tau\;
\hat{\K}(\tau){\rm e}^{-{\rm i}\omega_n\tau}=
\Ke{\int\frac{{\rm d}k}{2\pi}\frac{v_{\rm F}}
{\omega_n^2+\omega^2(k)}}^{-1}
\end{equation}
with $\:\omega_n=2\pi n/\beta\:$. Analytic continuation,
\begin{equation}\label{imteil}
J(\omega)=-\Im{\rm m}\lim_{\delta\to 0}\:\K(-{\rm i}\omega+\delta)
\end{equation}
relates $\:J(\omega)\:$ directly to $\:\K(\omega_n)\:$
\cite{weiss93}.

The asymptotic behaviours $\:J(\omega\to 0)\approx
2|\omega|/g\:$ and $\:J(\omega\to\infty)\sim 2\omega\:$ can
readily be deduced from (\ref{kdef}) in view of
$\:{\omega^2(|k|\ll\alpha)}\approx v_{\rm F}^2k^2/g^2\:$ and
$\:\omega^2(|k|\gg\alpha)\approx v_{\rm F}^2k^2\:$. Here,
$\:g\equiv(1+v_0/\pi v_{\rm F})^{-1/2}\:$ has been defined in
analogy to the Luttinger model with $\:v_0=\hat{v}(k=0)\:$
(cf.\ (\ref{fourier})). For $\:\omega(k)\:$ as in
(\ref{dispersion}) the integration (\ref{kdef}) with
(\ref{imteil}) can be carried out analytically, yielding
\begin{equation}\label{analytical}
\frac{J(\tilde{\omega})}{\alpha v_{\rm F}}=2\sqrt{2}\tilde{\omega}
\frac{( N_{+}(\tilde{\omega})+\tilde{\omega} N_{-}(\tilde{\omega}))
M(\tilde{\omega})}{(1+\tilde{\omega}^2)(|N_{+}(\tilde{\omega})|^2+
|N_{-}(\tilde{\omega})|^2)}
\end{equation}
where
\nnumb
M(\tilde{\omega})&=&\d\sqrt{4\tilde{\omega}^2+
(\tilde{\omega}^2-1/g^2)^2}\\[3ex]
N_{+}(\tilde{\omega})&=&\d\left|-\tilde{\omega}^2+
1/g^2+M(\tilde{\omega})\right|^{1/2}\\[3ex]
N_{-}(\tilde{\omega})&=&\d\left|-\tilde{\omega}^2+
1/g^2-M(\tilde{\omega})\right|^{1/2}\quad.
\nnume
The time $\:(\alpha v_{\rm F})^{-1}\:$ for electrons of velocity
$\:v_{\rm F}\:$ needed to traverse the interaction range establishes
the natural frequency scale of the problem,
$\:\tilde{\omega}=\omega/\alpha v_{\rm F}\:$.

Figure~1 illustrates the result (\ref{analytical}). At
small $\:\omega\ll\alpha v_{\rm F}\:$, $\:J(\omega)\sim 2\omega/g\:$
and the current-voltage relation (\ref{iv}) is recovered at
long wave lengths and low energies.

With large $\:\omega\gg\alpha v_{\rm F}\:$, $\:J(\omega)\:$
approaches the linear behaviour, $\:J(\omega)\to 2\omega\:$,
that corresponds to the non-interacting case, $\:g=1\:$, for the
reasons motivated initially. Note, however that $\:J(\omega)\:$
does not simply interpolate between either of the Ohmic
asymptotics but crosses the value $\:2\omega\:$ so that the
damping $\:J(\omega\gapprox \alpha v_{\rm F})<2\omega\:$ is {\it
smaller} than it would be in the absence of interactions.

\begin{figure}
\vspace*{7.6cm}\includegraphics{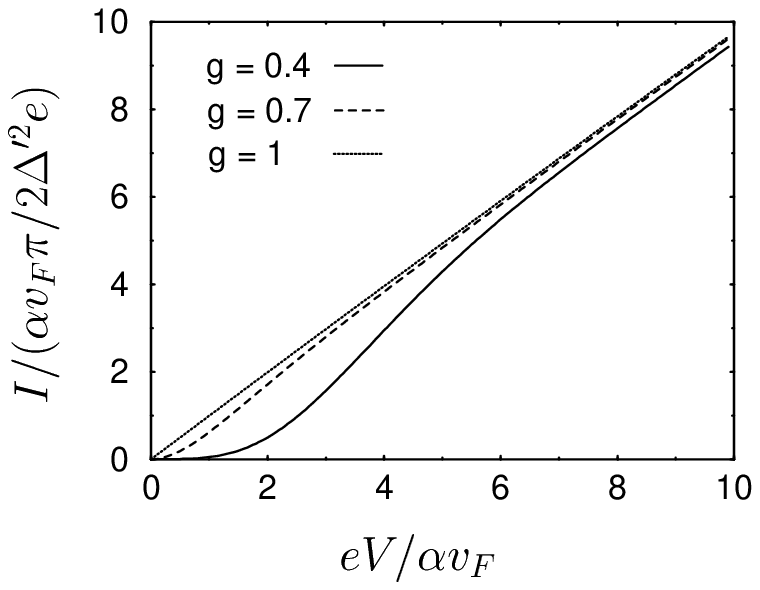}
\vspace{0.5cm}
\fcap{Figure~2: Current through the tunnel barrier versus
applied voltage for different $\:g\:$,
$\:\Delta'\equiv\Delta/\omega_c\:$. The power-law at
low voltages agrees with the Luttinger liquid behaviour. The
crossover to the linear relation, $\:I(V)=V/R_{\rm T}\:$,
manifests the finite range of the $\:e-e$--interaction.}
\end{figure}

The current-voltage relation, obtained according to
(\ref{iv-final}) for zero temperature, is depicted in Figure~2.
The crossover behaviour of $\:J(\omega)\:$ shows up in a
transition from the power law at low voltages, $\:I(V)\sim
V^{2/g-1}\:$ to the linear tunnel resistance behaviour,
$\:I(V)\to V/R_{\rm T}\:$. It takes place on the voltage scale
$\:\alpha v_{\rm F}(2/g-1)/e\:$. The high voltage limit does
{\em not} show any offset that would correspond to a Coulomb
blockade since the charging energy
\begin{equation}\label{charging}
E_c/e=\lim_{V\to\infty}\Kr{V/R_{\rm T}-I(V)}\propto
\int_0^{\infty}{\rm d}\omega\;\Kr{\frac{J(\omega)}{\omega}-2}=0
\end{equation}
vanishes. The proportionality in (\ref{charging}) follows from
the short time behaviour of $\:w(t)\:$ to the order $\:\sim
t^2\:$ (cf.\ \ref{instint}) and the right hand side expression
vanishes since, for any dispersion (\ref{dispersion}),
$\:\int{\rm d}\omega\;J(\omega)/2\omega\:$ equals the number of
modes in the wire, cf.\ (\ref{kdef},\ref{imteil}). Two
conditions are usually considered as being sufficient to
establish a Coulomb blockade \cite{devoret90,schoen90}~: the
suppression of quantum fluctuations of the charges by low
transmittivities and the presence of a nearby dissipative
environment of high impedance for which the bulk modes serve
\cite{sassetti94}. Although both conditions are fulfilled in
the present system no charging effects appear. The vanishing
lateral extension of a 1D wire does not suffice to accumulate
charging energy. Near a single barrier, a finite cross section
is required for the capacitance $\:C\:$ to be finite so that
$\:I(V)\sim (V-E_c/e)/R_{\rm T}\:$ at high voltages, with
$\:E_c=e^2/2C\:$ \cite{egger97}. This is consistent with the
result for a selfconsistent determination of the charge
distribution along the wire \cite{egger96}.

Eq.~(\ref{charging}) holds for any interaction potential of
finite range. Another example is the screened Coulomb
interaction $\:v(x-x')=e^2{\rm e}^{-
\alpha|x-x'|}/\linebreak\kappa\sqrt{(x-x')^2+d^2}\:$
($\:\kappa\:$: dielectric constant, $\:d\:$: width of the wire)
which is again determined by two parameters, $\:\alpha\:$ and
$\:\alpha v_0/2\to e^2/\kappa d\:$. At low voltages $\:I(V)\:$
obeys a power law and at $\:V>2\alpha v_{\rm
F}(\sqrt{1+2e^2K_0(\alpha d)/\kappa\pi v_{\rm F}}-1)/e\:$
($\:K_0\:$: modified Bessel function) a crossover to the
effectively non-interacting behaviour occurs. Only true
long-range interaction $\:\alpha\to 0\:$ changes the power-law
behaviour at low voltages \cite{gogolin94a,schulz93b,glazman92a}
and the divergence of $\:\hat{v}(k\to 0)\:$ suppresses the
crossover.

In principle, we can also infer the current-voltage
characteristics for the case of a weak barrier and attractive,
finite--range interactions by taking advantage of the exact
duality relation \cite{fisherzwerger} between the weak and the
strong barrier limit as it has been proven \cite{sassetti96} for
arbitrary $\:J(\omega)\:$. In our case, the dual
$\:J(\omega)\:$ interpolates between $\:J(\omega\ll\alpha v_{\rm
F})=2g\omega\:$ and $\:J(\omega\gg\alpha v_{\rm F})=2\omega\:$,
since $\:g\:$ maps to $\:1/g\:$, so that the interaction
strength again vanishes at high energies. Correspondingly, 
to second order in the barrier height, the current exhibits
a crossover from the Luttinger liquid behaviour
$\:I(V)\sim(1-c(U_b^2)V^{2g-2})V\:$ at small voltages to the
Ohmic behaviour, $\:I(V)\sim(1-c(U_b^2))V\:$, at large voltages.

\begin{figure}
\vspace*{15cm}\includegraphics{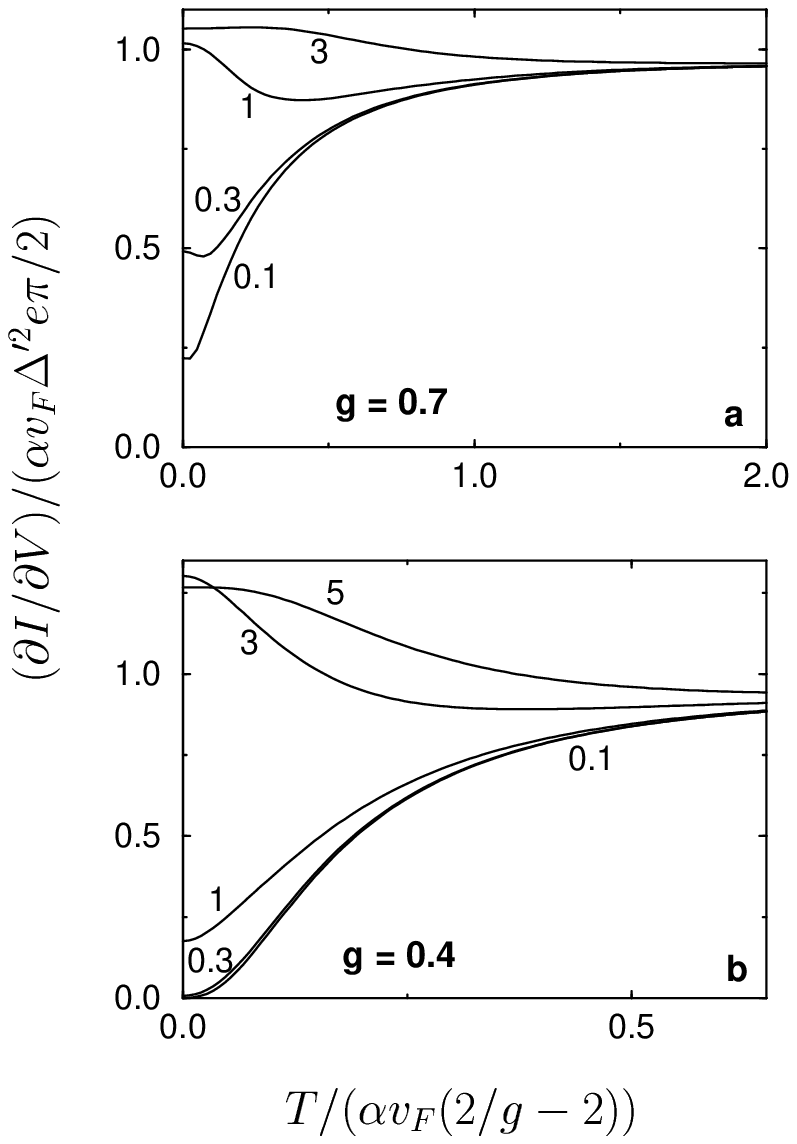}
\vspace{0.5cm}
\fcap{Figure~3: Temperature dependence of the differential
conductance for different voltages $\:eV/\alpha v_{\rm F}=
0.1,\ldots,5\:$ as indicated. In {\sf a} and {\sf b}
$\:g=0.7\:$ and $\:g=0.4\:$, respectively. At low
temperatures and low voltages the power law behaviours
\cite{kanefisher} are recovered.}
\end{figure}

Also the temperature dependence of the differential conductance
$\:\partial I/\partial V\:$, Figure~3, reveals a crossover
around $\:T\approx\alpha v_{\rm F}(2/g-2)\:$ as can be deduced
from a high temperature expansion up to the cubic term in
(\ref{instint}) and (\ref{iv-final}). At small $\:T\:$ the
linear conductance varies $\:\sim T^{2/g-2}\:$, in agreement
with \cite{kanefisher}, while the non-linear conductance can
even decrease with temperature. At high temperatures
$\:\partial I/\partial V\:$ approaches the constant value
$\:1/R_{\rm T}\:$, irrespective of the voltage, as has been
demonstrated already for weak interaction \cite{glazman94a}.

To summarise, we have shown that finite ranges of the
$\:e-e$--interaction change the non-linear current through a
tunnelling barrier in a 1D wire qualitatively, compared to the
simple power-law behaviour. The latter is usually considered to
be the main characteristic for one-dimensionality, but is only
valid for short range interactions. At high voltages the
influence of the interaction disappears and no Coulomb blockade
remains. Similarly, the differential conductance becomes
independent of high temperatures and assumes the value
$\:1/R_{\rm T}\:$ for all voltages.

Careful determination of the current voltage characteristics and
also of its temperature dependence would allow to measure
directly the range of the electron--electron interaction. This
quantity is difficult to access by other experimental means.

\noindent{\bf Acknowledgement}\\
We would like to thank H.~Grabert, B.~Kramer, M.~Sassetti, U.~Weiss,
and particularly R.~Egger for many valuable discussions and the
I.S.I.\ foundation for the fruitful atmosphere that stimulated
the present work under contract ERBCHRX-CT920020. MS acknowledges
support by the Gottlieb-Daimler- and Karl-Benz-Stiftung.
\begin{raggedright}

\end{raggedright}

\end{document}